 \documentclass[aps,prb,twocolumn,groupedaddress, amsmath, amssymb, showpacs]{revtex4-1}
\newcommand{\SP}{\textrm{SP}}

\newcommand{\vc}{\mathbf}
\usepackage{graphicx}
\usepackage{color}

\begin{document}


\title{Hall magnetohydrodynamic reconnection in the plasmoid unstable regime}


\author{S.\ D.\ Baalrud}
\author{A.\ Bhattacharjee}
\author{Y.-M.\ Huang}
\author{K.\ Germaschewski}

\affiliation{Center for Integrated Computation and Analysis of Reconnection and Turbulence, University of New Hampshire, Durham, New Hampshire 03824, USA}

\date{\today}

\begin{abstract}

A set of reduced Hall magnetohydrodynamic (MHD) equations are used to evaluate the stability of large aspect ratio current sheets to the formation of plasmoids (secondary islands). Reconnection is driven by resistivity in this analysis, which occurs at the resistive skin depth $d_\eta \equiv S_L^{-1/2} \sqrt{L v_A/\gamma}$, where $S_L$ is the Lundquist number, $L$ the length of the current sheet, $v_A$ the Alfv\'{e}n speed, and $\gamma$ the growth rate. Modifications to a recent resistive MHD analysis [N.\ F.\ Loureiro, A.\ A.\ Schekochihin, and S.\ C. Cowley, Phys.\ Plasmas {\bf 14}, 100703 (2007)] arise when collisions are sufficiently weak that $d_\eta$ is shorter than the ion skin depth $d_i \equiv c/\omega_{pi}$. Secondary islands grow faster in this Hall MHD regime: the maximum growth rate scales as $(d_i/L)^{6/13} S_L^{7/13} v_A/L$ and the number of plasmoids as $(d_i/L)^{1/13} S_L^{11/26}$, compared to $S_L^{1/4} v_A/L$ and $S^{3/8}$, respectively, in resistive MHD. 

\end{abstract}




\maketitle


\section{Introduction}

The primary focus of modern magnetic reconnection research has been to explain why observed reconnection rates in high Lundquist number plasmas are much faster than predicted by the seminal Sweet-Parker theory.\cite{swee:58,park:63} Important events such as the eruption of flares in the solar corona ($S_L \gtrsim 10^{12}$) and the sawtooth collapse in magnetic fusion experiments ($S_L \gtrsim 10^{6}$) occur much faster than the resistive diffusion timescale predicted by the classical models.\cite{bhat:04,yama:10} Here $S_L = 4\pi L v_A/(c^2 \eta)$ is the Lundquist number based on the system size $L$, $v_A$ is the Alfv\'{e}n speed, and $\eta$ the resistivity. The discrepancy between observed and predicted timescales has led to the point of view that fast reconnection cannot occur within the framework of resistive magnetohydrodynamics (MHD), which the Sweet-Parker theory is based on. 

Recent work has uncovered a fundamental flaw in the Sweet-Parker model when applied to high-Lundquist-number plasmas.\cite{lour:07,samt:09,bhat:09,cass:09,huan:10,ni:10,uzde:10,huan:11,uzde:10} These studies have shown that the plasmoid instability, i.e., secondary island instability, which is excited when $S_L \gtrsim 10^4$, can significantly enhance the reconnection rate. In plasmoid dominated reconnection, the primary mechanism of reconnecting field lines is the formation of magnetic islands (flux ropes in 3D\cite{daug:11}). These super-Alfv\'{e}nic instabilities quickly grow, nonlinearly saturate, or coalesce, and convect out of the current sheet, carrying magnetic flux with them.  Copious ejection of flux ropes have been observed at reconnection sites in solar flares\cite{sava:10} as well as the Earth's magnetopause\cite{russ:78} and magnetotail\cite{chen:07}.

Although secondary islands have been studied for a long time,\cite{furt:63} only recently have the scaling properties of the most unstable mode been established for a Sweet-Parker current sheet.\cite{lour:07} The primary insight has been to account for the Lundquist number scaling of the current sheet width: $\delta_{\SP} = LS_L^{-1/2}$. In conventional tearing mode theory, the current sheet width is taken to be constant, and the subsequent growth rate scales as $S_L$ to a negative exponent: $S_L^{-3/5}$ or $S_L^{-1/3}$, for the constant-$\psi$ and nonconstant-$\psi$ regimes, respectively.\cite{copp:76} Accounting for the crucial feature that a Sweet-Parker layer becomes increasingly singular at high $S_L$, the classical dispersion relation for tearing modes shows that the growth rate of the most unstable plasmoid scales as $S_L^{1/4}$.\cite{bhat:09} Thinning of the current sheet at high $S_L$ plays a critical role, leading to the surprising result that the plasmoid instability becomes increasingly unstable at higher $S_L$. 

These resistive MHD studies have provided a convincing argument for the importance of plasmoids at high $S_L$, but resistive MHD is not valid when $S_L$ is too high.  Two common scenarios by which the resistive MHD approximation breaks down are: (1) the current sheet width becomes shorter than the ion skin depth $d_i = c/\omega_{pi}$, (2) the resistive skin depth for the plasmoid $d_\eta = \delta_{\SP} \sqrt{(v_A/L)/\gamma}$ becomes shorter than the ion skin depth. Here $\gamma$ is the plasmoid growth rate. The resistive skin depth is the length scale at which plasmoids form and, thus, magnetic field lines reconnect.  Scenario (1) is what is typically considered Hall reconnection. If the current-sheet width is smaller than $d_i$, a large aspect ratio current sheet (with Y-points) is no longer a viable equilibrium; it is typically (but not always, see Ref.~\onlinecite{huan:11}) replaced by an X-point geometry, with an additional out of plane quadrupole field. Reconnection proceeds much faster in the X-point configuration.\cite{huan:11,shep:10} Starting from a Sweet-Parker current sheet, plasmoids can cause a cascade to the $d_i$ scale because the current layer between plasmoids scales as $\delta \sim \delta_{\SP}/\sqrt{N}$ where $N$ is the number of plasmoids.\cite{daug:09} 

In this work, we consider scenario (2). Since the plasmoid instability is super-Alfv\'{e}nic, $\gamma \gg v_A/L$, the resistive skin depth is much shorter than the current sheet width: $d_\eta/\delta_{\SP} = \sqrt{(v_A/L)/\gamma} \ll 1$. Thus, a Sweet-Parker current sheet that is thinning because of plasmoid production, or increasing $S_L$, will always enter a regime where Hall effects alter the plasmoid instability [scenario (2)] before the current sheet becomes thin enough to alter the equilibrium [scenario (1)]. This intermediate region, where $d_\eta \ll d_i \ll \delta_{\SP}$, provides a transition between plasmoid dominated reconnection in resistive MHD, where $d_i \ll d_\eta \ll \delta_{\SP}$, and conventional Hall reconnection, where $\delta_{\SP} \ll d_i$. It is important to check that plasmoids continue to be unstable in this intermediate regime, otherwise it could prevent the cascade to shorter scales. However, we find that this is not a concern because plasmoids are formed even more copiously in the Hall-plasmoid regime: the maximum growth rate scales as $\gamma_\textrm{H}/\Gamma_o \simeq \delta_i^{6/13} S_L^{7/13}$ and the number of plasmoids as $N_\textrm{H} \simeq \delta_i^{1/13} S_L^{11/26}$, compared to $\gamma_\textrm{R}/\Gamma_o \simeq S_L^{1/4}$ and $N_\textrm{R} \simeq S^{3/8}$, respectively, in resistive MHD.\cite{lour:07} Here $\delta_i \equiv d_i/L$ and $\Gamma_o \equiv 2v_A/L$.

\begin{figure}
\includegraphics{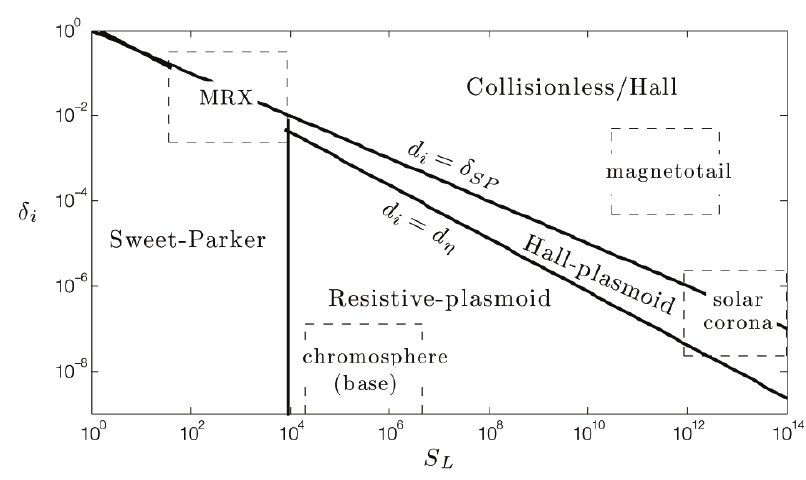}
\caption{Diagram of different reconnection scenarios in a $\delta_i$--$S_L$ space. Also shown are the parameter range for several plasmas of interest: the magnetic reconnection experiment (MRX), Earth's magnetotail, the base of the solar chromosphere, and the solar corona.}
\label{fg:valid}
\end{figure}

The Hall-plasmoid region we are interested in is defined by the boundaries $d_\eta \lesssim d_i \lesssim \delta_{\SP}$, which implies $S_L^{-5/8} \lesssim \delta_i \lesssim S_L^{-1/2}$. In this last expression, either $\gamma_\textrm{R}$ or $\gamma_\textrm{H}$ can be used in $d_\eta$ since both give $\gamma/\Gamma_o \simeq S_L^{1/4}$ at $d_i = d_\eta$ for the fastest growing mode. Figure~\ref{fg:valid} shows these boundaries in $\delta_i$--$S_L$ space.  Plasmas to the left of the vertical line at $S_L=10^4$ are considered stable to plasmoids.\cite{ni:10} This result is based on numerical simulations which find that the plasmoid growth rate is sub-Alfv\'{e}nic for $S_L \lesssim 10^4$, so any islands convect out of the reconnection layer before growing to significant amplitudes. A large aspect ratio current sheet (Y-points) is often not a viable equilibrium for plasmas above the $d_i = \delta_{\SP}$ line. Plasmas below the $d_i = d_\eta$ line are in the conventional resistive MHD regime. The Hall-plasmoid region we are concerned with is a narrow part of the $\delta_i$--$S_L$ parameter space at small $S_L$, but expands to cover a larger range of $\delta_i$ for high-$S_L$ plasmas such as the solar corona. Even for lower $S_L$, this region is important because it is always traversed as plasmoids cause a cascade from the resistive MHD regime to the Hall regime at shorter scales. 

Section~\ref{sec:rhe} describes the reduced Hall MHD equations, and Sec.~\ref{sec:equilib} the equilibrium, that are used in the linear tearing mode analysis in Sec.~\ref{sec:lgr}. The stability analysis is carried out by asymptotically matching solutions in the three layers (i) $x \ll d_i$, (ii) $d_\eta \ll x \ll \delta_{\SP}$, and (iii) $x \gg d_i$. It is shown that the basic equations describing the resistivity-driven Hall reconnection are similar to a collisionless reconnection problem solved by Mirnov, Hegna, and Prager in Ref.~\onlinecite{mirn:04}, and Fitzpatrick and Porcelli in Ref.~\onlinecite{fitz:04}. The physics is different because magnetic field lines reconnect due to resistivity at the $d_\eta$ scale here, instead of electron inertia at the $d_e$ scale in the collisionless problem. Nevertheless, the same boundary layer analysis can be used here. The properties of a Hall-plasmoid reconnection scenario are discussed in Sec.~\ref{sec:hplas}. The analytic results are compared with linear and nonlinear simulations in Sec.~\ref{sec:num}. Section~\ref{sec:num} also shows numerical solutions for the eigenfunctions, including the quadrupole out-of-plane magnetic field that arrises in the Hall-plasmoid regime.  Results are summarized in Sec.~\ref{sec:sum}.

\section{Reduced Hall MHD equations\label{sec:rhe}}

In terms of the normalized variables
\begin{align}
\hat{\vc{x}} = \frac{\vc{x}}{a} \ &, \ \ \hat{\vc{V}} \equiv \frac{\vc{V}}{v_A} \ , \ \hat{t} \equiv \frac{v_{A} t}{a} , \\ \nonumber
\hat{P} \equiv \frac{P}{B_o^2/(4\pi)} \ &, \ \hat{\vc{B}} \equiv \frac{\vc{B}}{B_o} \ , \ \hat{\vc{J}} \equiv \frac{\vc{J}}{cB_o/(4\pi a)} , \\ \nonumber
\hat{\vc{E}} \equiv \frac{\vc{E}}{v_A B_o/c} \ &, \ \hat{d}_i \equiv \frac{c/\omega_{pi}}{a} \ , \ S \equiv \frac{4\pi a v_A}{c^2 \eta} , \nonumber
\end{align}
the single fluid equation of motion and the generalized Ohm's law can be written 
\begin{equation}
(\partial_t + \vc{V} \cdot \nabla) \vc{V} = \vc{J} \times \vc{B} - \nabla P , \label{eq:mom}
\end{equation}
and
\begin{equation}
\vc{E} + \vc{V} \times \vc{B} = S^{-1} \vc{J} + d_i ( \vc{J} \times \vc{B} - \nabla P) ,\label{eq:ohm}
\end{equation}
respectively. The hats on normalized variables in Eqs.~(\ref{eq:mom}) and (\ref{eq:ohm}) have been omitted for notational convenience. Normalized variables will be used throughout Secs.~\ref{sec:rhe}, \ref{sec:equilib} and \ref{sec:lgr}.  Here $v_A \equiv B_o/\sqrt{4\pi \rho}$ is the Alfv\'{e}n speed and $\omega_{pi} \equiv \sqrt{4\pi e^2 n/m_i}$ is the ion plasma frequency. We will also use the Maxwell equations $\nabla \cdot \vc{B} = 0$, $\nabla \times \vc{E} = - \partial_t \vc{B}$, and $\nabla \times \vc{B} = \vc{J}$. Equations~(\ref{eq:mom}) and (\ref{eq:ohm}) assume that pressure can be represented by a scalar, and Eq.~(\ref{eq:ohm}) neglects electron inertial scale physics ($d_e \rightarrow 0$). We assume that $S^{-1} \gg d_e$, so reconnection is driven by resistivity. It has been shown recently that in collisionless ($S^{-1} \ll d_e$) and weakly collisional ($S^{-1} \sim d_e$) regimes, more general tensor descriptions of pressure can be required.\cite{hoss:09,fitz:10}

We assume incompressible flow, $\nabla \cdot \vc{V} =0$, and introduce the stream function $\phi$,
\begin{equation}
\vc{V} = \nabla \phi \times \hat{z} + V_z \hat{z} ,
\end{equation}
and flux function $\psi$,
\begin{equation}
\vc{B} = \nabla \psi \times \hat{z} + B_z \hat{z}  .
\end{equation} 
In terms of the stream and flux functions, Eqs.~(\ref{eq:mom}) and (\ref{eq:ohm}) can be written as the following set of four reduced resistive-Hall MHD equations:\cite{crai:05}
\begin{equation}
\partial_t V_z = [\phi, V_z] + [B_z, \psi] + V_o , \label{eq:rv}
\end{equation}
\begin{equation}
\partial_t \nabla^2 \phi = [\phi, \nabla^2 \phi] + [\nabla^2 \psi, \psi] ,  \label{eq:rphi}
\end{equation}
\begin{equation}
\partial_t \psi = S^{-1} \nabla^2 \psi + [\phi, \psi] + d_i[\psi, B_z] + E_o , \label{eq:rpsi}
\end{equation}
\begin{equation}
\partial_t B_z = S^{-1} \nabla^2 B_z + [\phi, B_z] + [V_z, \psi] + d_i [ \nabla^2 \psi, \psi]  . \label{eq:rb}
\end{equation} 
Equations~(\ref{eq:rv}) and (\ref{eq:rphi}) are obtained from the curl of Eq.~(\ref{eq:mom}) in the perpendicular (to $\hat{z}$) and parallel directions, respectively.  Likewise, Eqs.~(\ref{eq:rpsi}) and (\ref{eq:rb}) are obtained from the curl of Eq.~(\ref{eq:ohm}). Here $V_o$ and $E_o$ are constants, and $[f,g] = (\nabla f \times \nabla g) \cdot \hat{z}$ is the Poisson bracket. Similar reduced Hall MHD equations are also derivable from gyrokinetics, if a strong guide field is present.\cite{sche:09}

\section{Equilibrium\label{sec:equilib}}

To compare with the resistive MHD analysis of Loureiro \textit{et al.},\cite{lour:07} we want to use the same Sweet-Parker equilibrium configuration. Although we are using Hall MHD equations here, it is still possible to construct an equilibrium that does not depend on the ion inertial scale terms because the current sheet thickness is much larger than $d_i$. We choose the same linear flow profile as Louriero: $V_{xo} = - \Gamma_o x$, $V_{yo} = \Gamma_o y$ inside the current sheet ($-x_o < x < x_o$), $V_{xo} = -V_o$, $V_{yo}=0$ above the current sheet ($x > x_o$), and $V_{x} = V_o$, $V_{yo} =0$ below the current sheet ($x<-x_o$) . Here $\Gamma_o = 2 a/L$ in normalized units ($\Gamma_o = 2v_A/L$ in dimensional units). The associated stream function profile is
\begin{eqnarray}
\phi_o = \label{eq:estream} 
\left\lbrace \begin{array}{ll}
-\Gamma_o xy, & |x| \leq x_o \\
-\Gamma_o x_o y, & x > x_o \\
\Gamma_o x_o y, & x < - x_o
\end{array} \right. .
\end{eqnarray}

\begin{figure}
\includegraphics{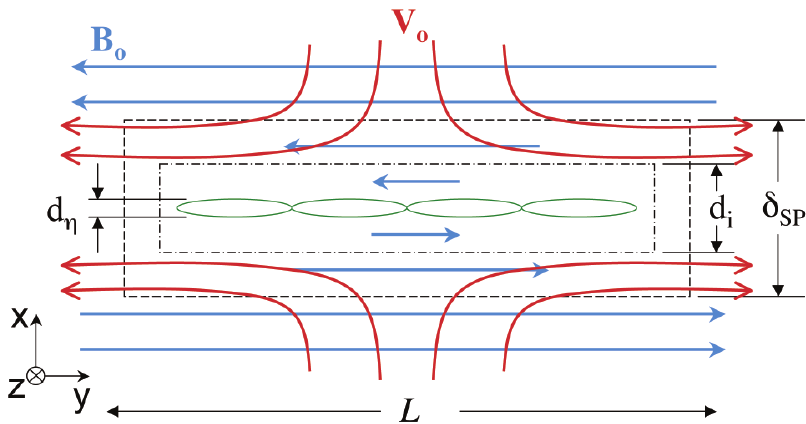}
\caption{Schematic drawing of three important scales in the Hall-plasmoid analysis: $d_\eta$, $d_i$, and $\delta_{\SP}$. The sheared equilibrium magnetic field, $\vc{B}_o$, and flow, $\vc{V}_o$, profiles are also shown.}
\label{fg:draw}
\end{figure}

Assuming $V_{zo}$ and $B_{zo}$ are constant, and looking for a steady-state 1D solution of the form $\psi_o = \psi_o(x)$, the only non-trival reduced equation is Eq.~(\ref{eq:rpsi}), which is
\begin{equation}
\frac{1}{\Gamma_o S} \frac{d B_{yo}}{dx} + x B_{yo} =\frac{E_o}{\Gamma_o} , \label{eq:byde}
\end{equation}
for $|x| \leq x_o$. Taking $B_{yo}(x=0) = 0$ as the boundary condition, and matching the solution of Eq.~(\ref{eq:byde}) to $B_{yo} = \pm 1$ for $|x|> \pm x_o$, yields Louriero's 1D equilibrium\cite{lour:07}
\begin{eqnarray}
B_{yo} = \label{eq:byo} 
\left\lbrace \begin{array}{cl}
\alpha \exp(-x^2/\hat{\delta}_{\SP}^2) \frac{\sqrt{\pi}}{2} \textrm{erfi} (x/\hat{\delta}_{\SP}) , & |x| \leq x_o \\
1,  & x > x_o \\
-1, & x < - x_o 
\end{array} \right. .
\end{eqnarray}
in which $\textrm{erfi}$ is the imaginary error function. Here $\hat{\delta}_{\SP} = \delta_{\SP}/a$ is the normalized Sweet-Parker width. The transition point $x_o$ is determined from the maximum of $B_{yo}$, which gives $x_o = 0.92 \hat{\delta}_{\SP}$ and $\alpha = 1.85$.\cite{note2} 

The detailed functional form of Eq.~(\ref{eq:byo}) will not be important in the inner region of the linear tearing mode analysis to follow. The critical feature is that the width of $B_{yo}$ is $\delta_{\SP}$. Thus, the current sheet becomes increasingly singular as the resistivity decreases. Equation~(\ref{eq:byo}) shows that $\delta_{\SP}$ is a convenient parameter to use to normalize length scales. We choose the length $a$, which has been arbitrary until now, to be 
\begin{equation}
a = \delta_{\SP} ,
\end{equation}
so that $\hat{\delta}_{\SP} = 1$. The schematic drawing of the sheared magnetic field and flow profiles is shown in Fig.~\ref{fg:draw}, along with the three scales, $d_\eta$, $d_i$, and $d_\SP$, that will be important in the tearing mode analysis to follow. 

\section{Linear growth rate\label{sec:lgr}}


\subsection{Linearized equations}

We linearize Eqs.~(\ref{eq:rv})--(\ref{eq:rb}) according to
\begin{equation}
\psi (x,y,t) = \psi_o(x) + \psi_1(x) e^{iky + \gamma t} ,
\end{equation} 
with analogous definitions for $\phi$, $V_z$, and $B_z$. We assume that the instability growth rate is much faster than the timescale for equilibrium flows into or out of the layer: $\gamma \gg \partial_x \phi_o , \partial_y \phi_o \sim \Gamma_o$, i.e., it is super-Alfv\'{e}nic. In this case, the linear shear flow only contributes through $B_{yo}(x)$.\cite{lour:07} 

Applying these assumptions, the linearized version of Eqs.~(\ref{eq:rv})--(\ref{eq:rb}) in the reconnection layer reduce to:
\begin{equation}
\gamma V_{z1} = ik B_{yo} B_{z1}  , \label{eq:rv2}
\end{equation}
\begin{equation}
\gamma (\partial_x^2 - k^2) \phi_1 = ik[B_{yo} (\partial_x^2 - k^2) - B_{yo}^{\prime \prime} ] \psi_1 , \label{eq:rphi2}
\end{equation}
\begin{equation}
\gamma [1 - d_\eta^2 (\partial_x^2 - k^2)] \psi_1 = ik B_{yo} \phi_1 - d_i i k B_{yo} B_{z1} , \label{eq:rpsi2}
\end{equation} 
\begin{align}
\gamma [1 - d_\eta^2 (\partial_x^2 - k^2)] B_{z1} &= ik B_{yo} V_{z1} \label{eq:rb2} \\ \nonumber
&+ d_i ik [B_{yo}(\partial_x^2 - k^2) - B_{yo}^{\prime \prime}] \psi_1 . 
\end{align}
In resistive MHD, tearing mode behavior is determined from just two equations: those for the the stream and flux functions [(\ref{eq:rphi2}) and (\ref{eq:rpsi2})]. In Hall MHD, four equations must be solved because the out of plane velocity and magnetic field are important at spatial scales of order $d_i$.  

\subsection{Inner region}

In the inner region, $x \ll 1$ ($x \ll \delta_{\SP}$ in unnormalized units), so $\partial_x^2 \gg k^2$ and $B_{yo} \simeq \alpha x$. Here, Eqs.~(\ref{eq:rv2})--(\ref{eq:rb2}) reduce to: 
\begin{equation}
g ( 1 - d_\eta^2\, \partial_x^2 ) \psi_1 = i x \phi_1 - i d_i  x B_{z1} , \label{eq:rpsi3}
\end{equation}
\begin{equation}
g ( 1 - d_\eta^2\, \partial_x^2 + x^2/g^2 ) B_{z1} = i d_i x \partial_x^2 \psi_1 , \label{eq:rb3}
\end{equation}
\begin{equation}
g\, \partial_x^2 \phi_1 = i x \partial_x^2 \psi_1 , \label{eq:rphi3}
\end{equation}
in which $V_{z1}$ has been eliminated by putting Eq.~(\ref{eq:rv2}) into (\ref{eq:rb2}), and $g \equiv \gamma/(\alpha k)$. 

Equations~(\ref{eq:rpsi3})--(\ref{eq:rphi3}) are similar to Eqs.~(41)--(43) of the collisionless tearing mode problem solved by Fitzpatrick and Porcelli.\cite{fitz:04} In fact, making the substitutions $d_e \rightarrow d_\eta$ and $c_\beta = \sqrt{\beta/(1+\beta)} \rightarrow 1$ in Ref.~\onlinecite{fitz:04}, yields identically Eqs.~(\ref{eq:rpsi3})--(\ref{eq:rphi3}). Fitzpatrick and Porcelli solve their equations using boundary layer theory by splitting the inner region into two parts: the narrowest part on the electron inertial scale, $x \sim d_e \ll d_i$, and a broader part on the ion inertial scale, $d_e \ll (x \sim d_i) \ll 1$. A growth rate solution is obtained by matching these two layer equations, as well as the ion inertial layer equation to the ideal MHD solution at large $x$, 
\begin{equation}
\phi_1 \rightarrow \phi_o \biggl[1 + \frac{2}{\Delta^\prime} \frac{1}{x} + \mathcal{O} (x^{-2}) \biggr]
\end{equation}
where $\Delta^\prime$ is the tearing stability index\cite{furt:63}
\begin{equation}
\Delta^\prime \equiv \frac{1}{\psi_1(0)} \biggl[ \frac{d \psi_1}{dx} \biggl|_{0^+} - \frac{d \psi_1}{dx} \biggl|_{0^-} \biggr] .
\end{equation}
The primary expansion parameter in Ref.~\onlinecite{fitz:04} is $g/(c_\beta d_i) \ll 1$. 

Fitzpatrick and Porcelli's analysis can be carried over to solve the Hall MHD problem described by Eqs.~(\ref{eq:rpsi3})--(\ref{eq:rphi3}) as long as the orderings $d_\eta \ll d_i \ll 1$ (in normalized units) are obeyed.\cite{note} In this case, the two parts that the inner region is split into are: the narrowest part on the scale of the resistive skin depth, $x \sim d_\eta \ll d_i$, and a broader part on the ion inertial scale, $d_\eta \ll (x \sim d_i) \ll 1$. Here, the expansion parameter is $g/d_i \ll 1$, which will be checked a posteriori based on the growth rate solution. Following Ref.~\onlinecite{fitz:04}, the growth rate (in normalized units) is
\begin{equation}
-\frac{\pi}{\Delta^\prime} = \frac{\pi}{2} \frac{g^2}{d_i G(g/d_i)} - \frac{d_\eta d_i G(g/d_i)}{g} , \label{eq:growth1}
\end{equation}
in which
\begin{equation} 
G(x) \equiv \frac{\sqrt{x}}{2} \frac{\Gamma (1/4 + x/4)}{\Gamma (3/4 + x/4)} ,
\end{equation}
and $\Gamma (x)$ is the Gamma function. Since the growth rate analysis is based on $g/d_i \ll 1$, we will use 
\begin{equation}
G(g/d_i) \simeq \frac{1}{2} \frac{\Gamma(1/4)}{\Gamma(3/4)} \sqrt{\frac{g}{d_i}} . \label{eq:gapprox}
\end{equation}
The general form of Eq.~(\ref{eq:growth1}) was first obtained by Mirnov, Hegna, and Prager.\cite{mirn:04} Calculating the tearing mode growth rate from Eq.~(\ref{eq:growth1}) requires the tearing stability index, $\Delta^\prime$, which is determined by $\psi_1$ in the outer region.

\subsection{Outer region}

In the outer region $x \gg d_i$, the out of plane magnetic and velocity perturbations decouple from the flux and stream function equations because the $d_i$ term in Eq.~(\ref{eq:rpsi2}) is negligible. The $d_\eta$ term in Eq.~(\ref{eq:rpsi2}) is also small, so the outer region is an ideal MHD solution.  Here, Eq.~(\ref{eq:rpsi2}) reduces to $\phi_1 = - i \gamma/(kB_{yo}) \psi_1$, and Eq.~(\ref{eq:rphi2}) to 
\begin{equation}
\psi_1^{\prime \prime} = \biggl( \frac{B_{yo}^{\prime \prime}}{B_{yo}} + k^2 \biggr) \psi_1  . \label{eq:loupsi}
\end{equation}

Equation~(\ref{eq:loupsi}) describes the same outer region as the resistive analysis of Loureiro {\it et al},\cite{lour:07} who solved it perturbatively by matching solutions from the $|x|<x_o$ ($\partial_x^2 \gg k^2$) and $|x| > x_o$ ($\partial_x^2 \ll k^2$) asymptotic limits. The solution in the $|x| < x_o$ limit is 
\begin{equation}
\psi_1^\pm (x) = C_1^{\pm} B_{yo}(x) + C_2^{\pm} B_{yo}(x) \int_{\pm x_o}^{x} \frac{dz}{B_{yo}^2 (z)}  .
\end{equation}
Taking $B_{yo}$ to be continuous at $x=0$ and using $B_{yo} \simeq \alpha x$ for $x \ll 1$ yields $C_2^{\pm} = - \alpha \psi_1(0)$. The solution of Eq.~(\ref{eq:loupsi}) in the $|x| > x_o$ region is $\psi_1^\pm = C_3^\pm \exp(\mp kx)$. Matching the first derivative of these two solutions at $x=x_o$, using $B_{yo} (\pm x_o) = \pm 1$ and $B_{yo}^\prime (\pm x_o) = 0$ yields $C_3^\pm = \alpha \psi_1(0) \exp(kx_o)/k$. Matching the solutions themselves yields $C_1^\pm = \pm \alpha \psi_1(0)/k$. Thus, 
\begin{eqnarray}
\psi_1^{\pm} = \pm \frac{\alpha \psi_1(0)}{k} \label{eq:psiout} 
\left\lbrace \begin{array}{ll}
B_{yo}[1 \mp k \int_{\pm x_o}^x dz B_{yo}^{-2}] & |x| \leq x_o \\
\exp[ k (x_o \mp x)]  & |x| > x_o 
\end{array} \right. 
\end{eqnarray}
which yields the tearing stability index\cite{lour:07}
\begin{equation}
\Delta^\prime \simeq \frac{2 \alpha^2}{k}  . \label{eq:tsi}
\end{equation}

\section{Hall plasmoid reconnection\label{sec:hplas}}

\begin{figure}
\includegraphics{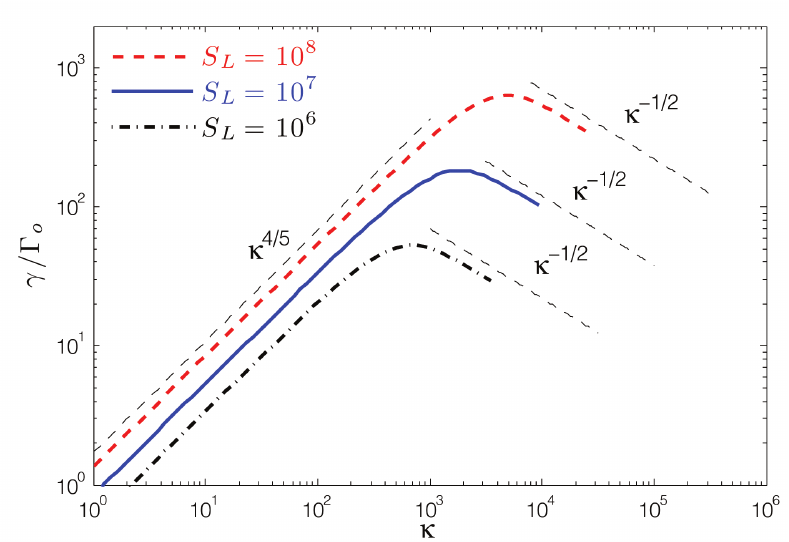}
\caption{Growth rate calculated from Eq.~(\ref{eq:gdeq}) using fixed ion skin depth $\delta_i = 10^{-4}$, and three values of the Lundquist number $S_L = 10^6, 10^7$ and $10^8$. }
\label{fg:gscale}
\end{figure}

\begin{figure}
\includegraphics{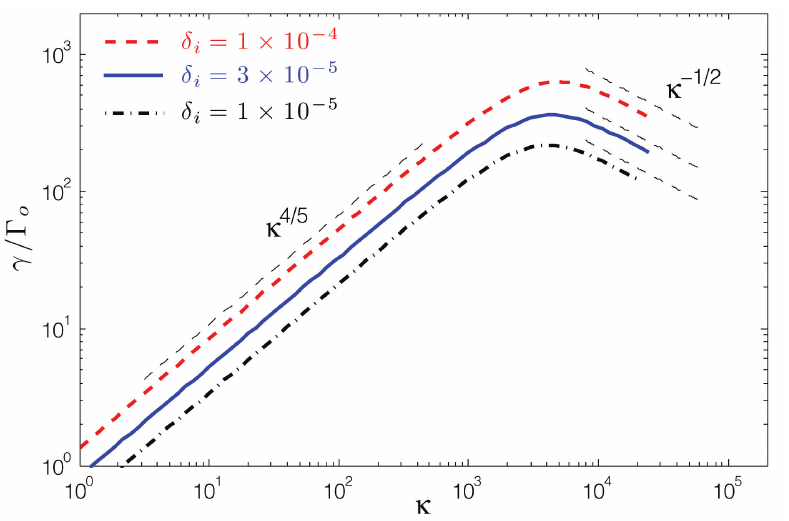}
\caption{Growth rate calculated from Eq.~(\ref{eq:gdeq}) using fixed Lundquist number $S_L = 10^8$, and three values of the ion skin depth $\delta_i = 1\times10^{-5}, 3\times10^{-5}$ and $1\times10^{-4}$. }
\label{fg:gdi}
\end{figure}

In terms of unnormalized, i.e., dimensional, variables, the dispersion relation from Eqs.~(\ref{eq:growth1}), (\ref{eq:gapprox}) and (\ref{eq:tsi}) is 
\begin{equation}
\biggl( \frac{\gamma}{\Gamma_o} \biggr)^{5/2} + \frac{c_g \kappa^{5/2} \delta_i^{1/2}}{(8\alpha)^{1/2} S_L^{1/4}} \frac{\gamma}{\Gamma_o} - \frac{\alpha^2 c_g^2 \kappa^2 \delta_i}{2\sqrt{\pi}} S_L^{1/2} = 0, \label{eq:gdeq}
\end{equation}
in which $\kappa \equiv kL$, $\delta_i \equiv d_i/L$, and
\begin{equation}
c_g \equiv \frac{1}{2} \frac{\Gamma (1/4)}{\Gamma (3/4)} \simeq 1.48 . 
\end{equation}
Figure~\ref{fg:gscale} shows $\gamma/\Gamma_o$ from a numerical solution of Eq.~(\ref{eq:gdeq}) for fixed $\delta_i = 1\times 10^{-4}$, and three values of Lundquist number, $S_L = 10^6, 10^7$ and $10^8$. Figure~\ref{fg:gdi} shows the solution for fixed $S_L = 10^8$, and three values of the ion skin depth $\delta_i = 1\times10^{-5}, 3\times 10^{-5}$ and $1 \times 10^{-4}$. 

Figures~\ref{fg:gscale} and \ref{fg:gdi} show that $\gamma/\Gamma_o$ has a power-law dependence with different slopes in the large and small wavenumber limits. For small $\kappa$, the second term in Eq.~(\ref{eq:gdeq}) is negligible. Here, the growth rate is given by
\begin{equation}
\frac{\gamma}{\Gamma_o} \simeq \frac{(\alpha c_g)^{4/5}}{(2 \sqrt{\pi})^{2/5}} \kappa^{4/5} \delta_i^{2/5} S_L^{1/5} . \label{eq:gsmall}
\end{equation}
For large $\kappa$, the first term in Eq.~(\ref{eq:gdeq}) is negligible. Here, the growth rate is given by
\begin{equation}
\frac{\gamma}{\Gamma_o} \simeq \sqrt{\frac{2}{\pi}} \alpha^{5/2} c_g \kappa^{-1/2} \delta_i^{1/2} S_L^{3/4} . \label{eq:glarge}
\end{equation}
The $\kappa^{4/5}$ and $\kappa^{-1/2}$ scalings predicted by Eqs.~(\ref{eq:gsmall}) and (\ref{eq:glarge}) match well with the numerical solutions of (\ref{eq:gdeq}). 

\begin{figure}
\includegraphics{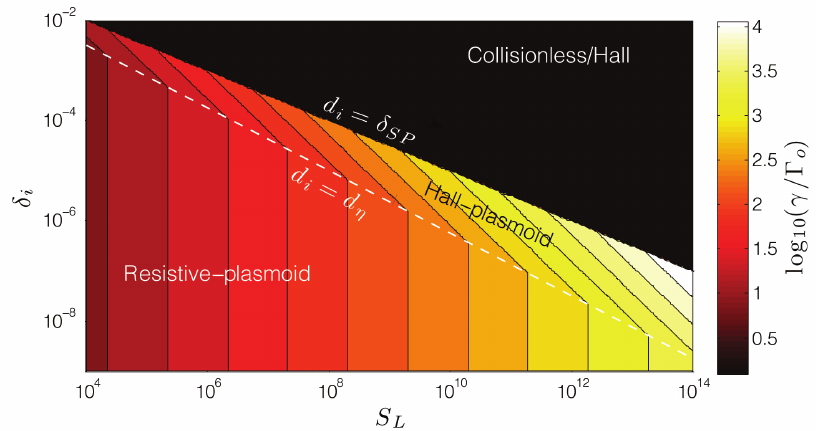}
\caption{Contour plot of the peak growth rate in the resistive-plasmoid ($\gamma/\Gamma_o \simeq S_L^{1/4}$) and Hall-plasmoid ($\gamma/\Gamma_o \simeq \delta_i^{6/13} S_L^{7/13}$) regions. }
\label{fg:gmax}
\end{figure}

Figures~\ref{fg:gscale} and \ref{fg:gdi} show that a broad range of wavenumbers can be unstable, but the most unstable will dominate the reconnection process. The wavenumber of the most unstable mode can be obtained by equating Eqs.~(\ref{eq:gsmall}) and (\ref{eq:glarge}), which yields 
\begin{equation}
\kappa_\textrm{max} \simeq 2.9\, \delta_i^{1/13} S_L^{11/26} \label{eq:kmax}
\end{equation}
Putting Eq.~(\ref{eq:kmax}) into either Eq.~(\ref{eq:gsmall}) or Eq.~(\ref{eq:glarge}) provides the growth rate of the most unstable plasmoid
\begin{equation}
\gamma_{\textrm{max}}/\Gamma_o \simeq 3.2\, \delta_i^{6/13} S_L^{7/13} . \label{eq:gmax}
\end{equation}
Since the number of plasmoids in the chain may be estimated to be $N \simeq \kappa_{\textrm{max}}/2\pi$, Eq.~(\ref{eq:kmax}) shows that the number of plasmoids formed for a given $\delta_i$ scales slightly more rapidly with Lundquist number in the Hall-MHD regime than the resistive-MHD regime (where $\kappa_{\textrm{max}} \simeq S_L^{3/8}$). The growth rate of the most unstable mode also grows more rapidly with $S_L$ ($\gamma_{\textrm{max}}/\Gamma_o \simeq S_L^{1/4}$ in the resistive case). Equation~(\ref{eq:kmax}) and Fig.~\ref{fg:gdi} show that $\kappa_{\max}$ is quite insensitive to the ion inertial length, being proportional to $\delta_i^{1/13}$. However, Eq.~(\ref{eq:kmax}) and Fig.~\ref{fg:gscale} show that $\kappa_{\max}$ is sensitive to Lundquist number. In contrast, the maximum growth rate is sensitive to both $\delta_i$ and $S_L$. Figure~\ref{fg:gmax} shows the peak growth rate as a function of $\delta_i$ and $S_L$ in the resistive-plasmoid and Hall-plasmoid unstable regions. The linear stability of plasmoids in the conventional Hall regime, $d_i \gtrsim \delta_{\SP}$, has yet to be worked out. 

Finally, we check the assumptions made during our analysis: $d_\eta \ll d_i \ll \delta_{\SP}$ and $g/d_i \ll 1$. The $d_\eta \ll d_i \ll \delta_{\SP}$ assumption requires $S_L^{-5/8} \lesssim \delta_i \lesssim S_L^{-1/2}$. For $\delta_i = 10^{-4}$, this implies $10^6 \lesssim S_L \lesssim 10^8$, which determines the $S_L$ limits in Fig.~\ref{fg:gscale}. Likewise, for $S_L=10^8$, this requires $10^{-5} \lesssim \delta_i \lesssim 10^{-4}$, which determines the $\delta_i$ limits in Fig.~\ref{fg:gdi}. The $g/d_i \ll 1$ (in normalized units) criterion can be written
\begin{equation}
\frac{2}{\alpha} \frac{S_L^{-1/2}}{\kappa \delta_i} \frac{\gamma}{\Gamma_o} \ll 1 . \label{eq:qcond}
\end{equation} 
For the most unstable mode, this reduces to $\delta_i \gtrsim S_L^{-5/8}$, which is the same requirement already obtained from the $d_\eta \ll d_i$ condition. Equation~(\ref{eq:qcond}) is more restrictive for $\kappa \neq \kappa_{\max}$, but we are primarily interested only in the most unstable mode here. Equation~(\ref{eq:tsi}) assumed $\kappa \ll S_L^{1/2}$, which sets the maximum $\kappa$ that can be considered in Figs.~\ref{fg:gscale} and \ref{fg:gdi}, but does not affect the most unstable mode (this will be discussed more in Sec.~\ref{sec:lin}). We also assumed that $\gamma/\Gamma_o \gg 1$, which sets the minimum $\kappa$ that can be considered in Figs.~\ref{fg:gscale} and \ref{fg:gdi}.

\section{Numerical Results\label{sec:num}}

\subsection{Numerical Linear Eigenmode Solutions\label{sec:lin}}

\begin{figure}
\includegraphics{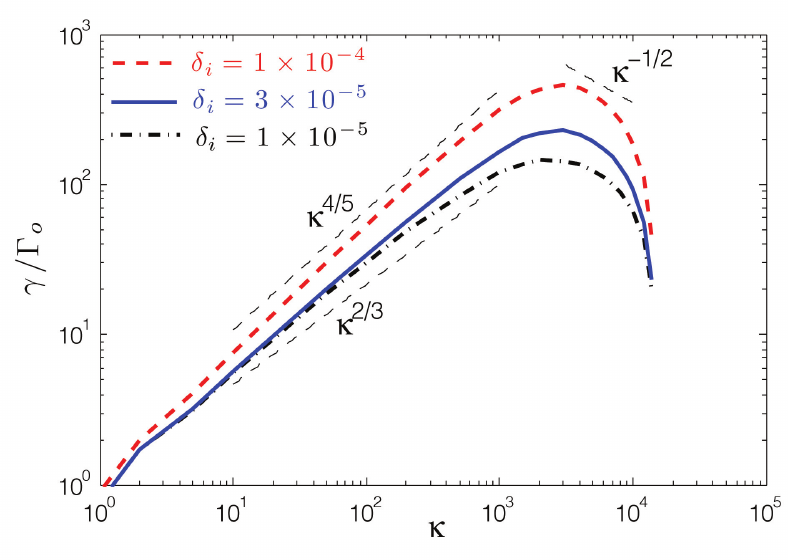}
\caption{Linear growth rate calculated numerically from Eqs.~(\ref{eq:rv2})--(\ref{eq:rb2}) using a fixed Lundquist number $S_L=10^8$, and three values of the ion skin depth $\delta_i = 1\times 10^{-5}$, $3\times 10^{-5}$, and $1\times 10^{-4}$.}
\label{fg:kai}
\end{figure}

The analysis of Sec.~\ref{sec:lgr} is based on the ordering $d_\eta \ll d_i \ll \delta_{\SP}$, but Fig.~\ref{fg:gmax} shows that the range of $d_i$ satisfying this can be narrow, especially at lower $S_L$. Thus, it is useful to check numerically the linear growth rate formulas obtained in Sec.~\ref{sec:hplas}. To do so, we solve numerically the four linearized reduced Hall MHD Eqs.~(\ref{eq:rv2})--(\ref{eq:rb2}). Figure~\ref{fg:kai} shows the plasmoid growth rate using a fixed Lundquist number $S_L=1\times 10^8$, and three values of the ion skin depth $\delta_i = 1\times 10^{-5}$, $3\times 10^{-5}$, and $1\times 10^{-4}$. These curves can be compared to the analytic results from Fig.~\ref{fg:gdi}. 

\begin{figure}
\includegraphics{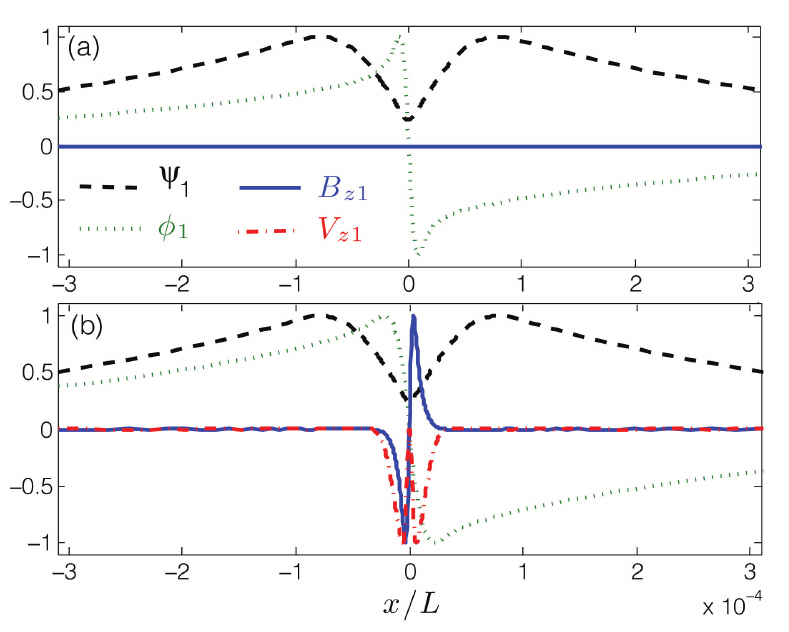}
\caption{Eigenfunctions for the perturbed fields $\psi_1$, $\phi_1$, $B_{z1}$, and $V_{z1}$ calculated numerically from Eqs.~(\ref{eq:rv2})--(\ref{eq:rb2}) using Lundquist number $S_L=10^8$, and two values of the ion skin depth: (a) a resistive MHD case with $\delta_i=0$, and (b) a Hall MHD case with $\delta_i = 3 \times 10^{-5}$. All eigenfunctions are normalized by their maximum amplitude.}
\label{fg:eig}
\end{figure}

For $S_L = 1\times 10^8$, the boundary between the resistive plasmoid and Hall-plasmoid regimes ($d_\eta = d_i$) occurs at $\delta_i \approx S_L^{-5/8} = 1\times 10^{-5}$. The boundary $d_i = \delta_{\SP}$ occurs at $\delta_i \approx S_L^{-1/2} = 1 \times 10^{-4}$. Figure~\ref{fg:kai} shows that for $d_i = 1 \times 10^{-5} = d_\eta$, the resistive MHD\cite{lour:07} scaling with $\kappa$, $\kappa^{2/3}$, is found. As $\delta_i$ increases through the Hall plasmoid regime, this steepens, in agreement with the theory. The Hall MHD scaling, $\kappa^{4/5}$, from Eq.~(\ref{eq:gsmall}), is found for the larger $\delta_i$.  

\begin{figure}
\includegraphics{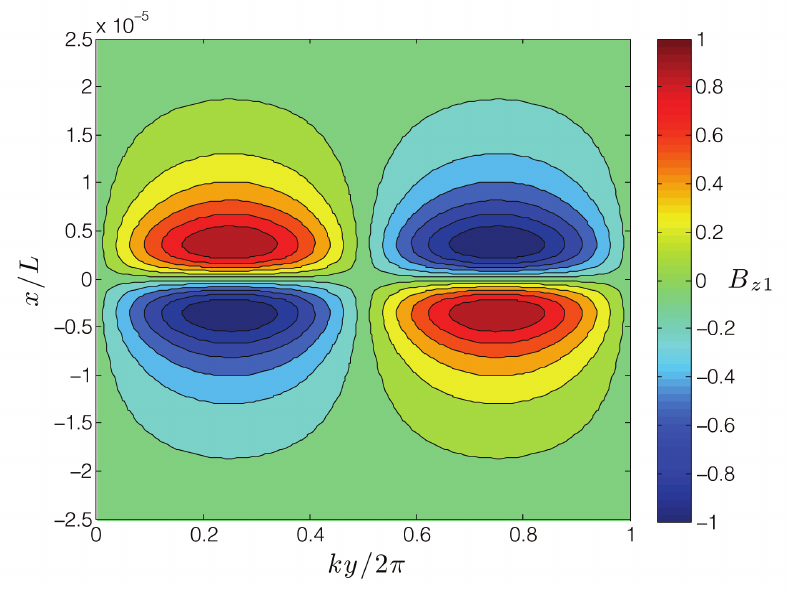}
\caption{Contour plot of constant $B_{z1}$ surfaces for a single period of the plasmoid chain. $B_{z1}$ has a quadrupole form, which is characteristic of the out-of-plane magnetic field in Hall reconnection. Here $S_L =  1\times10^8$, and $\delta_i = 3\times 10^{-5}$}
\label{fg:quad}
\end{figure}

The numerical results of Fig.~\ref{fg:kai} show that stabilization occurs quickly after the maximum $\kappa$ considered in the analytic theory is exceeded. The $\Delta^\prime$ solution used in the analytic analysis, Eq.~(\ref{eq:tsi}), assumed $k \delta_{\SP} \ll 1$ ($\kappa \ll S_L^{1/2}$). A more complete $\Delta^\prime$ solution that can account for this stabilization is $\Delta^\prime \approx 2\alpha^2 S_L^{1/2}/\kappa + 0.57 - 1.47 \kappa^2 S_L^{-1}$ (see Ref.~\onlinecite{lour:07}). Here we are primarily interested in the growth rate of the most unstable mode, which is accurately predicted by the approximate expression for $\Delta^\prime$. 

Figure~\ref{fg:eig} shows profiles of the eigenfunctions $\psi_1$, $\phi_1$, $B_{z1}$, and $V_{z1}$ for $S_L=10^8$ and two values of $\delta_i$. Each eigenfunction is normalized by its maximum amplitude. Panel (a) shows a resistive MHD solution, where the ion skin depth was taken to vanish $\delta_i = 0$. Here, the out of plane magnetic field and flow are zero, and the stream and flux functions of conventional MHD tearing modes\cite{furt:63,lour:07} are found. Panel (b) shows a case with $\delta_i = 3 \times 10^{-5}$, which, according to Fig.~\ref{fg:gmax}, is just into the Hall MHD regime ($d_\eta \lesssim d_i \lesssim \delta_{\SP}$) for this $S_L$. Figure~\ref{fg:eig} shows that the stream and flux functions have similar profiles in both the Hall and resistive MHD regimes, but that the out of plane magnetic field and flow velocity become important for spatial scales shorter than $d_i$ in Hall MHD. 

Figure~\ref{fg:quad} shows a contour plot of constant $B_{z1}$ surfaces over a single period of the plasmoid chain. The out-of-plane magnetic field generated by the presence of magnetic islands for this Hall MHD regime has a quadrupole form. Similar quadrupole fields have been observed in Hall MHD simulations of steady reconnection.\cite{tosh:83} However, these are typically seen when $d_i \gtrsim \delta_{\SP}$, and the equilibrium has an X-point configuration. Here quadrupole fields are formed due to the presence of plasmoids in an extended current-sheet equilibrium. 

\subsection{Evidence of the Instability in Nonlinear Simulations}

The prediction that Hall effects modify the plasmoid instability when $d_\eta \lesssim d_i$ can also be tested using more sophisticated codes that solve the full nonlinear Hall MHD equations. Here we employ the same simulation setup of two coalescing magnetic islands as in Refs.~\onlinecite{huan:10} and \onlinecite{huan:11}. Length scales are normalized to the length of the simulation box ($L=1$), which is square. Initially, the current sheet between the two islands is thicker than the Sweet-Parker width. As the simulation proceeds, the current sheet gradually becomes thinner until it reaches the Sweet-Parker width. To obtain the linear growth rate of plasmoids, we integrate $B_{x}^{2}$ at the central part of the current sheet along $x=0$, from $y=-1/4$ to $1/4$ at each time step. The magnitude of $f(t)\equiv \int_{-1/4}^{1/4}B_{x}^{2}(t)dy$ remains small before onset of the plasmoid instability, and increases abruptly after the onset. We fit $\ln(f)$ during the period that the plasmoid grows in accordance with $\ln(f) \simeq2\gamma t+c$. The $\gamma$ so obtained is a good measure of the linear growth rate of the fastest growing mode.

\begin{figure}
\includegraphics{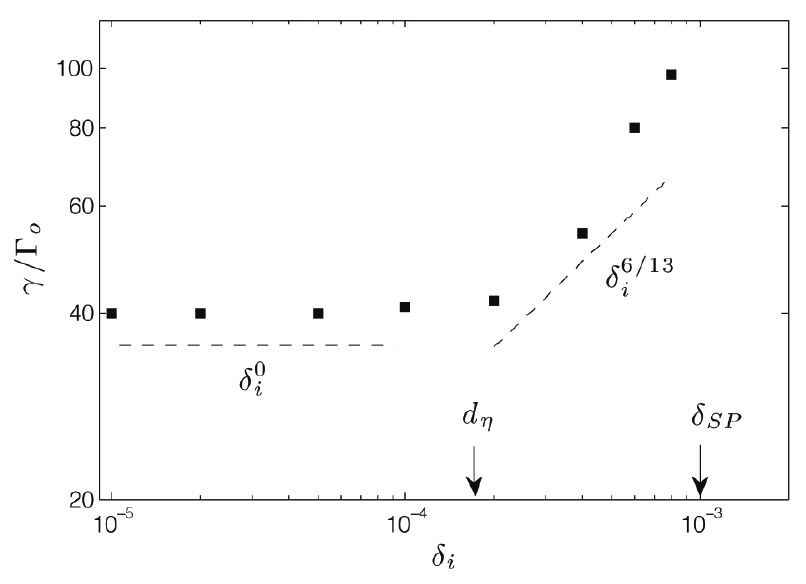}
\caption{Linear growth rate of plasmoids obtained from a numerical simulation of a full nonlinear set of Hall MHD equations. Here $S_L=1 \times 10^6$, for which $d_i=d_\eta$ when $\delta_i =1.8\times 10^{-4}$, and $d_i = \delta_{\SP}$ when $\delta_i = 1 \times 10^{-3}$. }
\label{fg:yimin}
\end{figure}

Figure~\ref{fg:yimin} shows the linear growth rate inferred from nonlinear Hall MHD simulations using $S_L=1\times 10^6$, and $\delta_i$ between $1\times 10^{-5}$ and $1\times 10^{-3}$. For $S_L = 1\times 10^6$, the Hall MHD regime is expected when $d_i \gtrsim d_\eta$, which implies $\delta_i \gtrsim S_L^{-5/8} = 1.8 \times 10^{-4}$. Figure~\ref{fg:yimin} shows that the transition between the resistive-plasmoid regime, where the growth rate is independent of $\delta_i$, and the Hall-plasmoid regime, where the growth rate depends on $\delta_i$, occurs around $\delta_i \simeq 2 \times 10^{-4}$. This agrees well with the analytic expectation. The analytic theory is not expected to hold for $\delta_i \gtrsim S_L^{-1/2} = 1\times 10^{-3}$, due to changes in the equilibrium. It is difficult to confirm the predicted $\delta_i^{6/13}$ scaling in the Hall-plasmoid regime because of uncertainties associated with inferring the linear growth rate from nonlinear simulations. However, confirming that Hall effects are important when the ion skin depth is shorter than $d_\eta$, rather than $\delta_{\SP}$, provides an important consistency check with the analytic theory. The Hall-plasmoid regime is expected to include a much broader range of $\delta_i$ for high-$S_L$ plasmas, such as the solar corona, but, unfortunately, computational resources limit the maximum $S_L$ obtainable in nonlinear simulations.

\section{Summary\label{sec:sum}}

There are two scale lengths that, when short enough, can cause the conventional resistive-MHD tearing mode theory to break down: the current sheet width ($\delta_{\SP}$), and the resistive skin depth ($d_\eta$). When $\delta_{\SP} \lesssim d_i$, the equilibrium magnetic configuration is expected to change. An X-point geometry is typically seen in this regime, which is often referred to as the Hall reconnection regime. However, if the plasmoid instability is present, the resistive skin depth is much shorter than the current sheet width: $d_\eta \ll \delta_{\SP}$. Thus, Hall effects modify the plasmoid instability properties at lower $S_L$ than is required to cause a change in the equilibrium configuration. As plasmoids cause a cascade to shorter scales, this means that the Hall-plasmoid regime is always realized before the scales become short enough to modify the equilibrium. Out of plane velocity and magnetic field components arise in the Hall-plasmoid regime, with the magnetic field having a quadrupole form. 

Plasmoids grow more rapidly with $S_L$ in the Hall-plasmoid regime than the resistive-plasmoid regime. The growth rate of the most unstable mode scales as $\delta_i^{6/13} S_L^{7/13}$ and the number of plasmoids as $\delta_i^{1/13} S_L^{11/26}$. The corresponding resistive-MHD scalings are $S_L^{1/4}$ and $S_L^{3/8}$. Thus, the cascade to shorter scales that the plasmoid instability causes is expected to continue through the Hall-plasmoid regime until the relevant current-sheet width (between plasmoids) becomes shorter than $d_i$. At this point, the Sweet-Parker equilibrium, characterized by an extended current-sheet, will be superseded by a Hall-MHD equilibrium.

\begin{acknowledgments}

The authors gratefully acknowledge conversations with Dr.\ Brian Sullivan and Dr.\ Will Fox. This research was supported in part by an appointment to the U.S. Department of Energy Fusion Energy Postdoctoral Research Program administered by the Oak Ridge Institute for Science and Education (S.D.B.), and DOE Grant No.\ DE-FG02-07ER46372, NSF Grant Nos.\ ATM-0802727, ATM-0903915, and AGS-0962698, and NASA Grant Nos. NNX09AJ86G and NNX10AC04G.

\end{acknowledgments}

\bibliography{refs.bib}

\begin{thebibliography}{99}
\bibitem{note2} The constant $\alpha$ is 1.31 in Ref.~\onlinecite{lour:07} because it uses $\delta_{\textrm{CS}} = \sqrt{c^2 \eta/(4\pi \Gamma_o)}$ to characterize the current sheet thickness. Here we use $\delta_{\SP}$, which is $\sqrt{2} \delta_{\textrm{CS}}$.
\bibitem{note} Parts of Ref.~\onlinecite{fitz:04} have been debated recently.\cite{fitz:07,hoss:09,fitz:10} However, this debate is concerned with determining if the electron pressure model used in \onlinecite{fitz:04} is appropriate for modeling collisionless tearing modes. This concern does not affect the Hall MHD problem we are interested in here.
\end{thebibliography}

\end{document}